\documentclass[10pt]{iopart}
\usepackage{graphicx} 
\usepackage{xcolor}

\begin{document}

\title[Detecting topological transitions]{Structural signatures of a topological transition in a disordered tetrahedral network}

\author{F Ellary$^1$, T C Nicholas$^2$, D Chakrabarti$^{3,4}$, A L Goodwin$^1$}
\address{$^1$Inorganic Chemistry Laboratory, University of Oxford, South Parks Road, Oxford OX1 3QR, U.K.}
\address{$^2$Center for Molecular Modeling (CMM), Ghent University, Technologiepark-Zwijnaarde 46, 9052 Zwijnaarde, Belgium}
\address{$^3$School of Chemistry, University of Birmingham, Edgbaston, Birmingham B15 2TT, UK}
\address{$^4$International Institute for Sustainability with Knotted Chiral Meta Matter (WPI-SKCM$^2$), Hiroshima University, 1-3-1 Kagamiyama, Higashi-Hiroshima, Hiroshima 739-8531, Japan}
\ead{andrew.goodwin@chem.ox.ac.uk}
\vspace{10pt}
\begin{indented}
\item[]\today
\end{indented}

\begin{abstract}
Topological transitions in disordered network materials are often characterised using descriptors of entanglement, such as ring linking and writhe, which can be difficult to access experimentally. Here we investigate whether a topological transition can instead be identified using structural descriptors more readily connected to experiment. We reanalyse configurations of a colloidal tetrahedral network model known to exhibit a pressure-driven transition between unentangled and and entangled states. We show that the cumulative all-rings vertex symbol (CARVS), a descriptor of medium-range network topology based on ring statistics, provides a clear signature of this transition despite containing no explicit information about ring linking. We further show that network entanglement produces characteristic changes in the radial distribution function and structure factor, including the emergence of short-range correlations between nonbonded nodes and the disappearance of a low-angle scattering pre-peak. These results establish experimentally accessible structural signatures of topological transitions in disordered tetrahedral networks. Comparison with canonical continuous random networks further suggests that framework materials with extended linkers may provide promising candidate systems in which to search experimentally for entanglement-driven topological transitions.
\end{abstract}

\section{Introduction}

\begin{figure}[t]
	\begin{center}
		\includegraphics[width=\columnwidth]{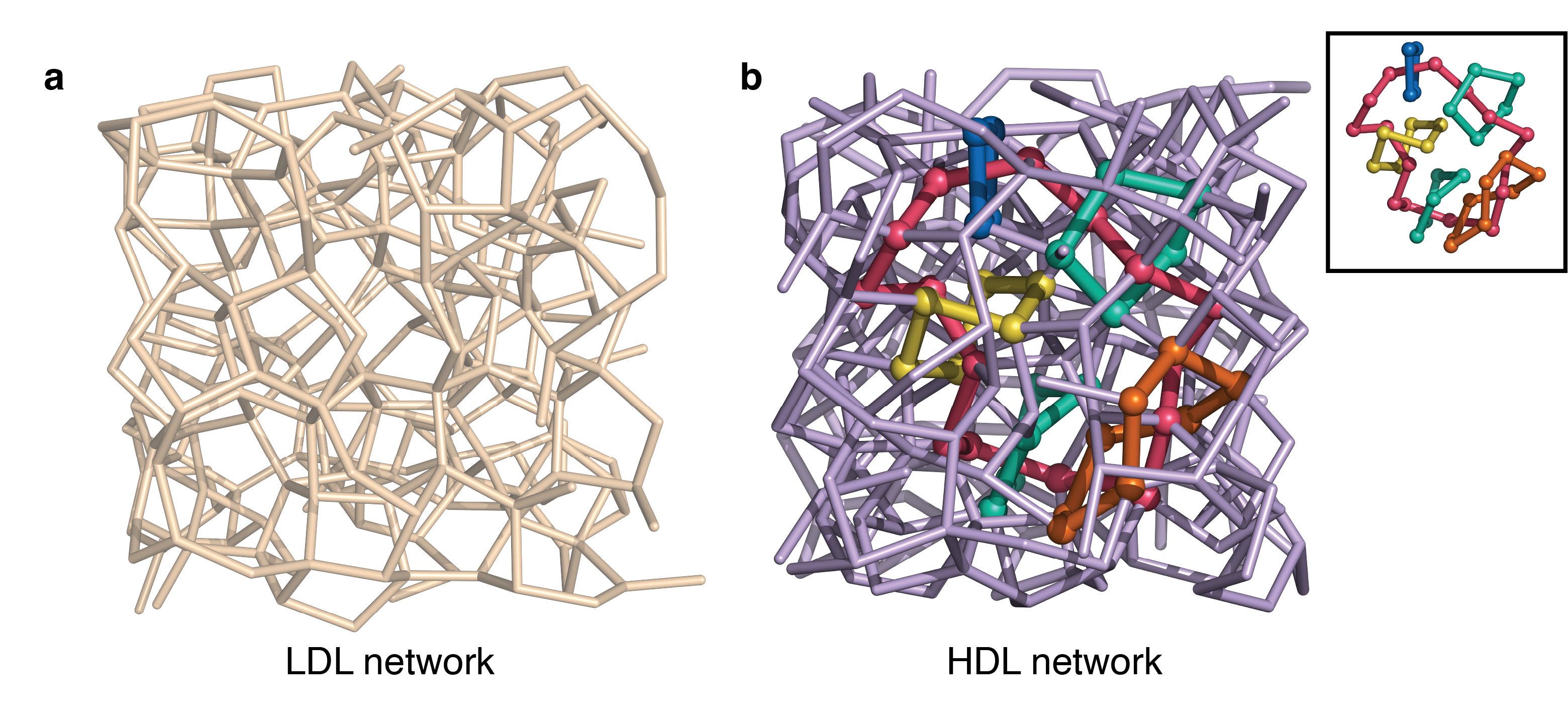}
	\end{center}
	\caption{\label{fig1} Network representation of typical configurations of the tetrahedral network formed by the colloidal model of water, conceived in Ref.~\cite{Neophytou2022}, on either side of a pressure-driven liquid--liquid phase transition it undergoes. In this representation, nodes are the centres of tetrahedral clusters formed of four triblock patchy particles and edges (\emph{i.e.} sticks) represent bonds between neighbouring tetrahedral clusters. (a) The low-density liquid (LDL) network is locally tetrahedral and topologically trivial: although it contains many rings, the rings are not linked. (b) The high-density liquid (HDL) retains predominantly tetrahedral local geometry but is substantially denser, and this densification is accompanied by the emergence of entanglement. Highlighted in colour (with nodes shown as spheres) is a 16-member ring (red) together with five smaller rings, each of which is linked with it; the inset shows the spatial embedding of these non-trivial topological motifs in isolation from the rest. The presence or absence of such linked motifs, as quantified by the network linking number, can distinguish the two phases.}
\end{figure}

Amorphous tetrahedral networks are known to exhibit phase transitions, marked by a change in network density \cite{Poole1992,Sastry2003,Hsu2008,Palmer2014,Debenedetti2020}. Remarkably, recent studies have unravelled a topological perspective of such phase transitions in model systems \cite{Neophytou2022,Neophytou2024,GutierrezFosado2024}, where the low- and high-density phases across the transitions are distinguished in terms of the absence or presence of entanglement, which in turn may correlate with physical characteristics such as diffusivity and/or elastic moduli [Fig.~\ref{fig1}]. A particularly insightful example---and the focus of this study---is the colloidal water model developed in Ref.~\cite{Neophytou2022} to interpret the proposed liquid--liquid phase transition in water \cite{Palmer2014,Debenedetti2020,Kim2017,Kim2020}. This model unambiguously shows a pressure-driven topological transition between two liquid phases, which can be continuous or discontinuous depending upon the temperature. While experimental evidence for the presence of a liquid-liquid phase transition in water is now available \cite{You2026}, a topological distinction between the two liquid networks is yet to be verified experimentally.   

The characteristic order parameters that track these transitions are topological descriptors, extracted through analysis of three-dimensional structural models \cite{Liu2021}. These descriptors include the linking number and writhe, neither of which has a definitive signature in experimental measurements. Hence a clear challenge with regards to the discovery of physical systems exhibiting topological transitions in disordered media is whether suitable topological descriptors can either be inferred robustly from other structural measures or indeed measured directly. 

Our own interest in the field is motivated by the increasing diversity of chemistries that are known to support disordered tetrahedral network structures. Examples include the porous aromatic frameworks that are a form of covalent organic framework \cite{Cai2022}, and the amorphous zeolitic imidazolate frameworks that are an important class of metal--organic frameworks \cite{Bennett2014}. Historically, the structures of such systems have often been interpreted in the context of continuous random network (CRN) models of \emph{e.g.}\ amorphous silicon \cite{Wooten1985,Bennett2010,Gaillac2017}, but more recently it has become possible to explore meaningfully the differences in ring statistics that occur between chemically different CRN systems \cite{Nicholas2025}. What is emerging is that increasing linker lengths seem to favour networks with larger and more diverse ring sizes, suggesting chemical routes to controlling the network topology of these heavily disordered materials. Hence the colloidal models developed initially to provide insight into the liquid--liquid transition in water may also find application in helping interpret the structures of these different classes of disordered tetrahedral network systems. In particular, appropriate chemistries may allow design of systems exhibiting \emph{bona fide} topological transitions under favourable conditions, such as ambient temperature and/or low hydrostatic pressures. Were such transitions to occur, it is important that we have strong and diverse experimental and computational tools to identify and characterise their nature.

There are two particular avenues that have piqued our interest and form the basis for the study we describe here. The first is a recently-developed ring-statistic metric, known as CARVS (= cumulative all-rings vertex symbol), which we have found to provide a richer and more sensitive descriptor of topological differences in amorphous networks \cite{Nicholas2025}. This descriptor and its derived functions contain no explicit information about ring linking, but are nonetheless capable of identifying key differences amongst various CRN topologies (\emph{e.g.}\ amorphous silicon \emph{vs} amorphous silica \emph{vs} amorphous zeolitic imidazolate frameworks). The second direction we wish to explore exploits the observation that entanglement can bring nodes that are distant in network connectivity into close proximity in real space, which should result in characteristic signatures in the radial distribution function (and hence the structure factor) \cite{Neophytou2024}. Signatures of entanglement in these functions are appealing because both the radial distribution function and the structure factor are experimentally accessible.

In this work, we therefore exploit the fact that the colloidal water model shows an unambiguous topological transition, and test whether this transition is measurable also using these alternative approaches. We reanalyse the same Monte Carlo configurations originally reported in Ref.~\cite{Neophytou2022}, calculating the CARVS descriptors, the radial distribution functions and the structure factors. Before reporting our results, we include a little relevant background to aid the reader: first, a brief recap of what is known regarding the transition itself; and, second, a pr\'{e}cis of the CARVS formalism (because they are not yet used widely) together with its relationship to established ring statistics. In the results section that follows, we apply the CARVS analysis to the colloidal water configurations, and show that it captures the topological transition in a manner that resembles closely the linking-number analysis presented by Neophytou {\it et al.} \cite{Neophytou2022}. We then proceed to calculate the corresponding pair correlation functions and structure factors, and find that these too are sensitive to entanglement in this particular model system. An advantage of the CARVS analysis we carry out is that it enables direct comparison between the colloidal water model and other canonical tetrahedral CRNs, and our results section concludes with a brief exploration of the similarities and differences amongst these systems. We conclude with a general discussion of the implications of our results for characterising topological transitions in amorphous materials using experimental structure factor measurements.

\section{Background}
\subsection{Colloidal water topological transition}

\begin{figure}
	\begin{center}
		\includegraphics[width=\columnwidth]{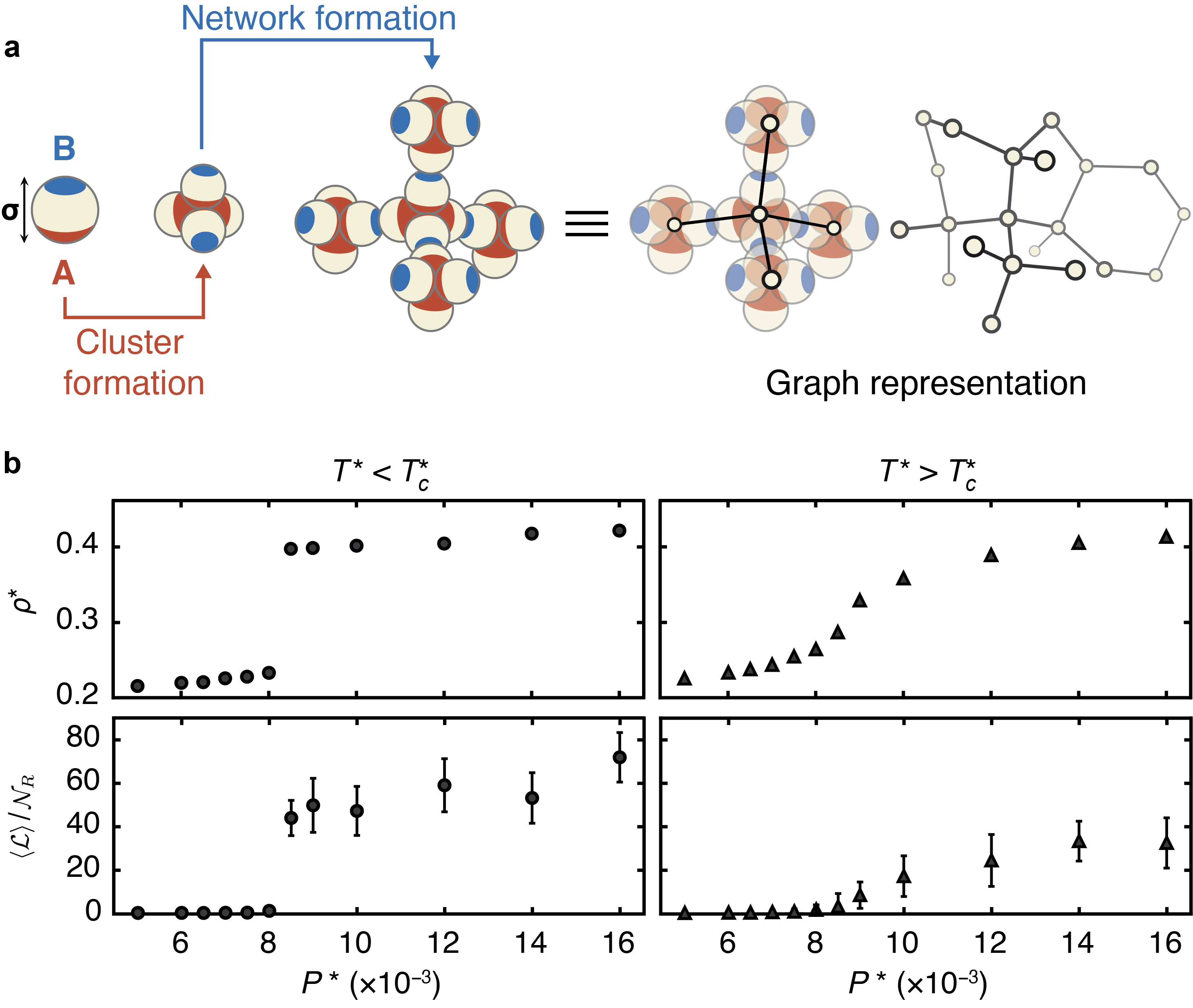}
	\end{center}
	\caption{\label{fig2}The colloidal water model and its topological transition, adapted from Ref.~\cite{Neophytou2022}. (a) Hierarachical self-assembly of triblock patchy particles of hard-core diameter $a$, each carrying an A patch (red) and a smaller B patch (blue) at opposite poles. Strongly attractive A--A interactions drive the formation of four-particle tetrahedral clusters, with the A patches buried in the interior and B patches directed outwards from its vertices. In the graph representation, each tetrahedral cluster is reduced to a single node at its centre and each B--B connection to an edge connecting nodes, yielding a predominantly four-fold connected network. (b) Reduced number density $\rho^\ast$ (top) and the average network linking number $\langle \mathcal L\rangle$ divided by the number of chordless cycles in the network $\mathcal N_{R}$ (bottom) as a function of reduced pressure $P^\ast$, at $T^\ast=0.1050>T_{\rm c}^\ast$ (left, circles) and $T^\ast=0.1120>T_{\rm c}^\ast$ (right, triangles).}
\end{figure}

We use this section to briefly describe the colloidal model of water central to our study and also summarise the key evidence for the existence of a topological transition in this model under pressure. The model involves colloidal triblock patchy particles, with a hard-core diameter of $a$, each containing two types of circular patches, labelled `A' and `B', on opposite poles across a middle band \cite{Neophytou2022}. Here, patches of the same type are attractive, but interactions between different types of patches are ruled out. Sufficiently strong A--A interactions relative to the B--B interactions drive a two-stage self-assembly into an amorphous network \emph{via} tetrahedra, formed by the interactions of four A patches buried in their interior. The smaller B patches point away from the vertices of these tetrahedra, leaving them with secondary valences. At a lower temperature, weaker B--B interactions drive the reversible self-assembly of a tetrahedral network structure, in which the vast majority of tetrahedra are connected \emph{via} B--B bonds [Fig.~\ref{fig2}(a)].

The thermodynamics of this model defines a critical point associated with the topological transition. Following Ref.~\cite{Neophytou2022}, energies are measured in units of the B--B interaction strength $\varepsilon_{BB}$, so that the reduced temperature, pressure, and number density are
\begin{eqnarray}
    T^\ast&=&k_{\rm B}T/\varepsilon_{BB}, \\
    P^\ast&=&Pa^3/\varepsilon_{BB},\\
    \rho^\ast&=&Na^3/V,
\end{eqnarray}
respectively.
Here, $a$ is the characteristic particle size, $N$ is the number of particles, and $V$ is the system volume. Using the same model parameters as in that work, the critical point lies at  $T_{\rm c}^\ast\simeq0.11$. At temperatures below this value, pressure drives a first-order transition from an LDL to an HDL phase; above $T_{\rm c}^\ast$, the corresponding structural and topological changes occur continuously [Fig.~\ref{fig2}(b)]. The topological nature of the transition is made evident by the calculation of the average network linking number $\langle \mathcal L\rangle$, which increases from zero in the LDL phase to a nonzero value in the HDL phase [Fig.~\ref{fig2}(c)]. The linking number between a pair of rings $R_i,R_j$ is a measure of how many times the two rings wind around each other, thus providing a metric for their mutual entanglement, and is given explicitly by the double line integral \cite{Liu2021}
\begin{equation}\label{lk}
\mathcal L(R_i,R_j)=\frac{1}{4\pi}\oint_{R_i}\oint_{R_j}\frac{\mathbf r_j-\mathbf r_i}{|\mathbf r_j-\mathbf r_i|^3}\cdot({\rm d}\mathbf r_j\times{\rm d}\mathbf r_i),
\end{equation}
computed over positions $\mathbf r_i,\mathbf r_j$ on the corresponding rings.

The interpretation of this transition in the colloidal water model is that the LDL phase can be described by a tetrahedral network with negligible ring linking; on application of pressure, there is a sudden increase in density (discontinuous if $T^\ast<T^\ast_{\textrm c}$ and continuous if $T^\ast>T^\ast_{\textrm c}$) associated with reorganisation into an interpenetrating, topologically-linked network. This is the process illustrated in Fig.~\ref{fig1}.

\subsection{CARVS}

\begin{figure}
	\begin{center}
		\includegraphics{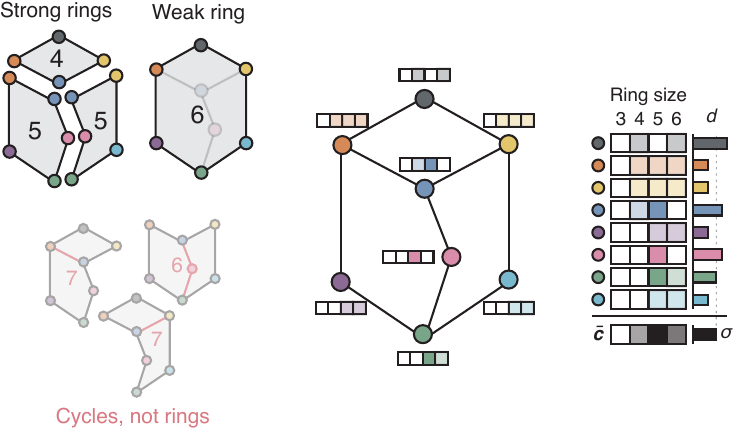}
	\end{center}
	\caption{\label{fig3}Schematic illustration of how the CARVS descriptors are obtained, using an eight-node network as an example. The four-membered ring and the two five-membered rings highlighted in the top left of the figure are \emph{strong} rings: none can be written as the sum of any number of smaller cycles. The six-membered cycle running around the outside of the network is a \emph{weak} ring: it is the sum of three smaller cycles (the strong rings), but not of two. Since we do not apply the strong-ring criterion, both strong and weak rings are counted. The six- and seven-membered cycles shown in the bottom left of the figure, by contrast, are not rings: each contains a shortcut (red), a path shorter than either route around the cycle between its end points. Equivalently, each is the sum of two smaller cycles. For each node $j$ (coloured circles), the number of rings of size $l$ containing that node gives the element $^la_j$ of the per-node vector $\mathbf c_j$, represented in the graphical table on the right of the figure, with darker shading for larger counts. Averaging over all nodes gives the network CARVS vector $\bar{\mathbf c}$ represented in grayscale. Coloured bars show the distance $d_j$ of the corresponding node vectors from $\bar{\mathbf c}$, and $\sigma$ is the root-mean-square of these distances, capturing the variance of node environments across the network.}
\end{figure}

The cumulative all-rings vertex symbol (CARVS) is a vector-valued descriptor of medium-range order in networks that can be used to characterise crystalline and noncrystalline nets alike \cite{Nicholas2025,Nicholas2026}. A CARVS vector is calculated for each node in a network structure, and a per-network CARVS vector is given by the average over these node vector terms. To determine the CARVS vector for a given node $j$, we enumerate all rings in the network structure that contain this node.\footnote{Following Refs.~\cite{DelgadoFriedrichs2005,Blatov2010}, we use the following definitions: an \emph{elementary cycle}---henceforth simply referred to as a \emph{cycle}---is a closed sequence of adjacent vertices in which no vertex, besides the coincident beginning and ending one, occurs more than once; the \emph{sum of two cycles} is the set of edges contained in exactly one of the cycles; and a \emph{ring} is a cycle that is not the sum of two smaller cycles (called a \emph{primitive ring} in Ref.~\cite{LeRoux2010}). In our analysis we include all rings, both \emph{strong} (not the sum of any number of smaller cycles) and \emph{weak} (all other rings).} There can be a very large number of rings containing any one node. In practice, the ring search is therefore carried out to increasing maximum size $l_{\texttt{max}}$ until the resulting CARVS vector is converged. For each parameter $^la_j$ we assign to it the number of rings of length $l$ that include node $j$. Then the CARVS for node $j$ is given by the vector
\begin{equation}
\mathbf c_j=\{^1a_j,^2a_j,\ldots,^La_j\},
\end{equation}
and the network CARVS vector is
\begin{equation}
\bar\mathbf c=\frac{1}{N}\sum^N_{j=1}\mathbf c_j,
\end{equation}
where $N$ is the number of nodes in a configuration [Fig.~\ref{fig3}].

For crystalline systems, there is a finite number of crystallographically distinct node environments, and the corresponding network CARVS therefore has a compact and highly constrained form, much like conventional vertex symbol descriptions of crystalline nets \cite{DelgadoFriedrichs2003,OKeeffe2008,Blatov2010}. Amorphous networks contain a much broader distribution of inequivalent node environments, and their averaged CARVS components are generally non-integer values extending over a wider range of ring sizes. By convention, when giving a CARVS vector for an amorphous system, only those elements with values $\geq1$ are generally reported (\emph{i.e.} where the average node has at least one ring of a given size).\footnote{This truncation is purely a reporting convention; all ring-size components are retained in the quantitative analyses to follow.} So, for example, it was shown in Ref.~\cite{Nicholas2025} that established models of amorphous silicon \cite{Deringer2018} are described by the CARVS $\{5_{2.0}.6_{5.2}.7_{4.1}\}$. The meaning of this symbol is that, on average, each Si atom in amorphous silicon is a member of 2.0 five-membered rings, 5.2 six-membered rings, and 4.1 seven-membered rings. Unlike conventional point and vertex symbols, which characterise only the shortest cycles and shortest rings, respectively, at each angle around a node, the CARVS includes all rings to which a node belongs. This gives the CARVS metric a heightened sensitivity to medium-range order relative to conventional topological descriptors \cite{Blatov2010,Nicholas2025}.

An additional, related, parameter that has proven useful in characterising topologies of disordered materials arises from the variance in node CARVS vectors for a given configuration. To quantify this measure we first calculate for each node its Euclidean distance
\begin{equation}
d_j=|\mathbf c_j-\bar\mathbf c|
\end{equation}
from the network average. We then define
\begin{equation}
\sigma=\sqrt{\frac{1}{N}\sum_{j=1}^Nd_j^2}
\end{equation}
which is the root-mean-squared Euclidean deviation of the node CARVS from the network average. It is large when the network contains a broad distribution of topologically distinct node environments, and small when those environments are comparatively uniform. The general trend identified in Ref.~\cite{Nicholas2025} for amorphous tetrahedral networks was that increasing the characteristic node--node separation resulted in an increasing population of larger rings and also an increased value of $\sigma$.

A key implication of our recent study of amorphous zeolitic imidazolate frameworks \cite{Nicholas2025} is that modern combinations of machine-learned interatomic potentials and high-quality total scattering data---for example as combined in hybrid reverse Monte Carlo methodologies---can in principle produce physically realistic structural models of amorphous network materials from experimental measurements. Here we use the term `realistic' to imply an accurate representation not only of local structure but also of medium-range network features such as described by the CARVS. If CARVS descriptors are sensitive to topological transitions, then experimentally-constrained structural models may identify such transitions even when the relevant topological information is not encoded explicitly in the scattering data themselves.

\section{Results and Discussion}
\subsection{CARVS analysis}

Given the fact that the colloidal water model of Ref.~\cite{Neophytou2022} exhibits a topological phase transition, our starting point was to determine the CARVS descriptors for representative configurations corresponding to relative pressure and temperature values on either side of this transition. We made use of the Monte Carlo configurations first reported in Ref.~\cite{Neophytou2022}, together with the corresponding adjacency lists used in that study to define network connectivity. We do not reconstruct the network using a distance cutoff, because interpenetration brings nodes that are distant in network connectivity into close spatial proximity. Our analysis focuses on two series of pressures along two isotherms: one at $T^\ast=0.1050<T^\ast_{\rm c}$ and the other at $T^\ast=0.1120>T^\ast_{\rm c}$. Each series included pressure points distributed over $0.005<P^\ast\leq0.016$, such that they spanned the topological transition. The $T^\ast=0.1120$, $P^\ast=0.0140$ state point was excluded because a substantial fraction of the available configurations showed evidence of crystallisation, which was independently apparent in the CARVS, radial distribution function, and structure factor analyses.

\begin{figure}
	\begin{center}
		\includegraphics[width=\columnwidth]{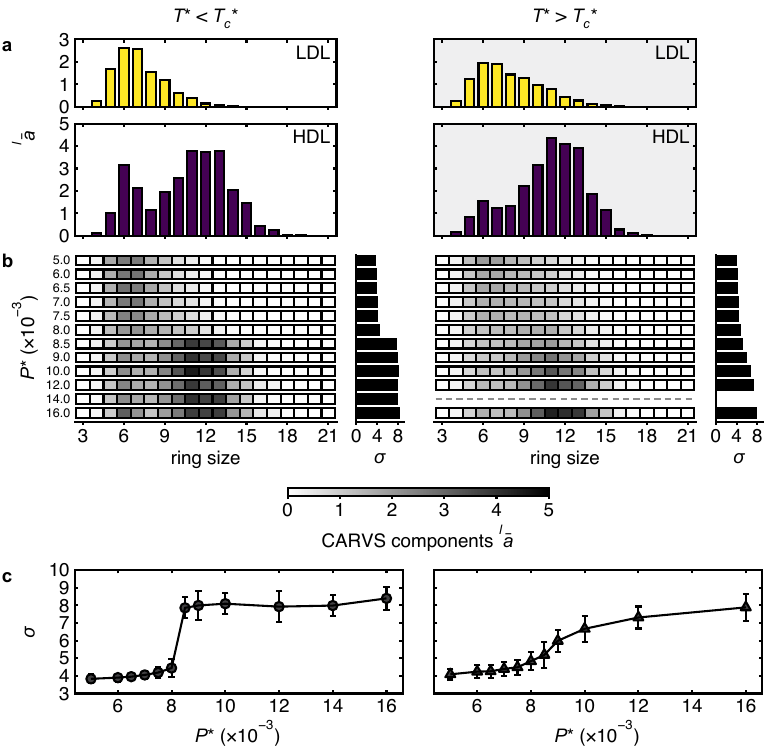}
	\end{center}
	\caption{\label{fig4}
    Pressure evolution of the CARVS descriptors below and above the critical temperature $T_{\rm c}^\ast$, shown in the left- and right-hand panels, respectively. 
    (a) Components $^l\bar a$ of the network CARVS vector $\bar{\mathbf c}$ as a function of ring size $l$, for representative LDL ($P^\ast=0.0050$, yellow) and HDL ($P^\ast=0.0160$, purple) configurations at each temperature. The LDL distribution is dominated by six- and seven- membered rings, whereas in the HDL phase a broad population of larger rings, of sizes $\approx10$--14, emerges.
    (b) CARVS barcodes for each pressure studied, with components shown on a greyscale, together with the corresponding values of $\sigma$. Below $T_{\rm c}^\ast$, the CARVS vector changes discontinuously from the LDL to the HDL distribution between $P^\ast=0.0080$ and $P^\ast=0.0085$; above $T_{\rm c}^\ast$, the change is continuous. The dashed line marks  $P^\ast=0.0140$ above $T_{\rm c}^\ast$, where both the CARVS and the structure factor show evidence of crysatllisation; this configuration is omitted. 
    (c) The CARVS distribution width $\sigma$ as a function of $P^\ast$, which likewise roughly doubles discontinuously below $T_{\rm c}^\ast$ and increases continuously above it. 
    }
\end{figure}

The evolution of the CARVS vectors for these state points along the two isotherms is shown in Fig.~\ref{fig4}(a). In both cases, the LDL CARVS is described by a single maximum at ring sizes of six or seven, followed by a gradual decrease in mean number of rings per node such that any rings larger than nine were present less frequently than one ring per node. As the topological transition is traversed under pressure, the form of this distribution is strongly modified. The HDL distributions are now bimodal, with maxima at ring sizes of around six and 12. Moreover, the total number of rings per node also increases substantially, together with both the width and the extent of the distributions. For example, even ring sizes of 14 are present with sufficient populations in which each node is a member of more than one, on average. These changes between the LDL and HDL distributions appear to be much more abrupt for the low-temperature isotherm than for the high-temperature isotherm.

We quantify the pressure-dependence of the CARVS variations at both temperatures by plotting the CARVS distribution width $\sigma$ as a function of pressure [Fig.~\ref{fig4}(c)]. This representation makes clear that the CARVS descriptor tracks an abrupt change between the LDL and HDL phases along the low-temperature isotherm, and a continuous evolution along the high-temperature isotherm. The qualitative similarity to the density and linking-number plots shown in Fig.~\ref{fig2}(b) and reported originally in Ref.~\cite{Neophytou2022} is striking. Because CARVS depends only on network connectivity and not on an absolute spatial length scale, the variation shown in Fig.~\ref{fig4}(c) cannot simply be a direct consequence of densification, as might be the case for changes in the positions of peaks in a pair correlation function. The close correspondence between CARVS and network linking number instead suggests that both capture the same underlying reorganisation of network topology.

\begin{figure}
	\begin{center}
		\includegraphics[width=\columnwidth]{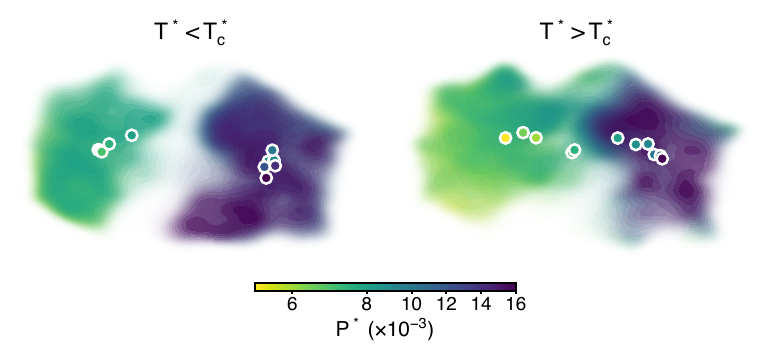}
	\end{center}
	\caption{\label{fig5}Two-dimensional UMAP embedding of the topological space spanned by the colloidal network configurations, below (left) and above (right) $T_{\rm c}^\ast$. Each underlying point corresponds to the CARVS vector of a single node in a single configuration. The shaded distributions show the density of these node-level points, coloured by the reduced pressure $P^\ast$ of the configuration from which they are drawn. Circles mark the projections of the network averaged CARVS vectors $\bar{\mathbf c}$ for each state point. A single embedding, fitted to the node-level CARVS vectors of all amorphous networks considered in this study, is used throughout, so that positions are directly comparable between panels. Below $T_{\rm c}^\ast$, LDL and HDL configurations occupy two disjoint regions of topological space, and the state-point averages jump discontinuously from one to the other across the trasnition. Above $T_{\rm c}^\ast$, the state-point averages traverse the same space continuously with increasing pressure, passing through the intermediate region that is unpopulated below $T_{\rm c}^\ast$.}
\end{figure}

The CARVS vectors allow us to represent visually a two-dimensional embedding of the topological space spanned by the various colloidal network structures included in our analysis. We show in Fig.~\ref{fig5} a projection of this topological distribution, generated using Uniform Manifold Approximation and Projection (UMAP) \cite{McInnes2018} for the low- and high-temperature isotherms. These distributions are generated by considering the points in high-dimensional space generated by the family of node-level CARVS vectors associated with all configurations. Each point in the UMAP representation therefore corresponds to a single node in a single configuration, while the larger markers show the state-point-average CARVS vectors.
The common UMAP embedding itself was fitted using the node-level CARVS vectors from all amorphous networks considered in this study, including those to be introduced later. In this embedding, nodes with similar topological environments end up projected onto nearby points in the UMAP representation. We think this projection provides a very appealing visual representation of the topological transition and its different natures at temperatures below and above the critical point. At low temperatures, the CARVS vectors are distributed entirely on the left-hand side of the UMAP projection at all pressures below the transition, and then there is a discontinuous jump to a different region of topological space at pressures above the transition (to the right-hand side of the projection). When analysing the high-temperature configurations using the same approach we see the same progression across topological space, but this progression is now much smoother and proceeds \emph{via} an intermediate region that is `forbidden' at temperatures below $T^\ast_{\rm c}$.

We therefore conclude that the CARVS descriptor is indirectly sensitive to the topological transition in this model. This is perhaps a surprising result because the linking number [Eq.~(\ref{lk})] explicitly encodes for topological entanglement, whereas the ring statistics from which the CARVS vectors are assembled contain no explicit information about ring linking. Our hypothesis is that the sensitivity of CARVS to entanglement arises from an increase in the number of large rings in order to allow interpenetration of tetrahedral networks [Fig.~\ref{fig1}]. The CARVS descriptor is particularly sensitive to these changes for two reasons. First, its node-centred construction effectively up-weights larger rings relative to conventional global ring-size distributions, because each ring contributes to the environments of all its constituent nodes.\footnote{For a network containing $N$ nodes and $R_l$ rings of size $l$, the network-averaged CARVS component is $^l \bar{a} = l R_l / N$. Thus, each ring is effectively weighted by its size, unlike in conventional global ring counts.} Second, whereas conventional point and vertex symbols characterise only the shortest cycles or rings associated individual node angles \cite{Blatov2010}, CARVS retains contributions from larger rings. The use of extended local ring environments has precedent in local-cluster analyses in \emph{e.g.} amorphous silicon \cite{Treacy2000,Treacy2012} and silica \cite{Hobbs1998}. By systematically incorporating these contributions into a quantitative descriptor, CARVS is sensitive to changes in the distribution of larger rings accompanying network entanglement, despite containing no explicit information about ring linking.

\subsection{Radial distribution function and structure factor}

We turn now to consider the evolution of the radial distribution function and its inverse Fourier transform, the structure factor, along the two isotherms studied here. Our motivation for investigating these functions was the observation that network interpenetration results in additional features in the radial distribution function at distances lower than the (bonded) nearest-neighbour peak, arising from correlations between non-bonded node pairs in close spatial proximity \cite{Neophytou2024}.

\begin{figure}
	\begin{center}
		\includegraphics{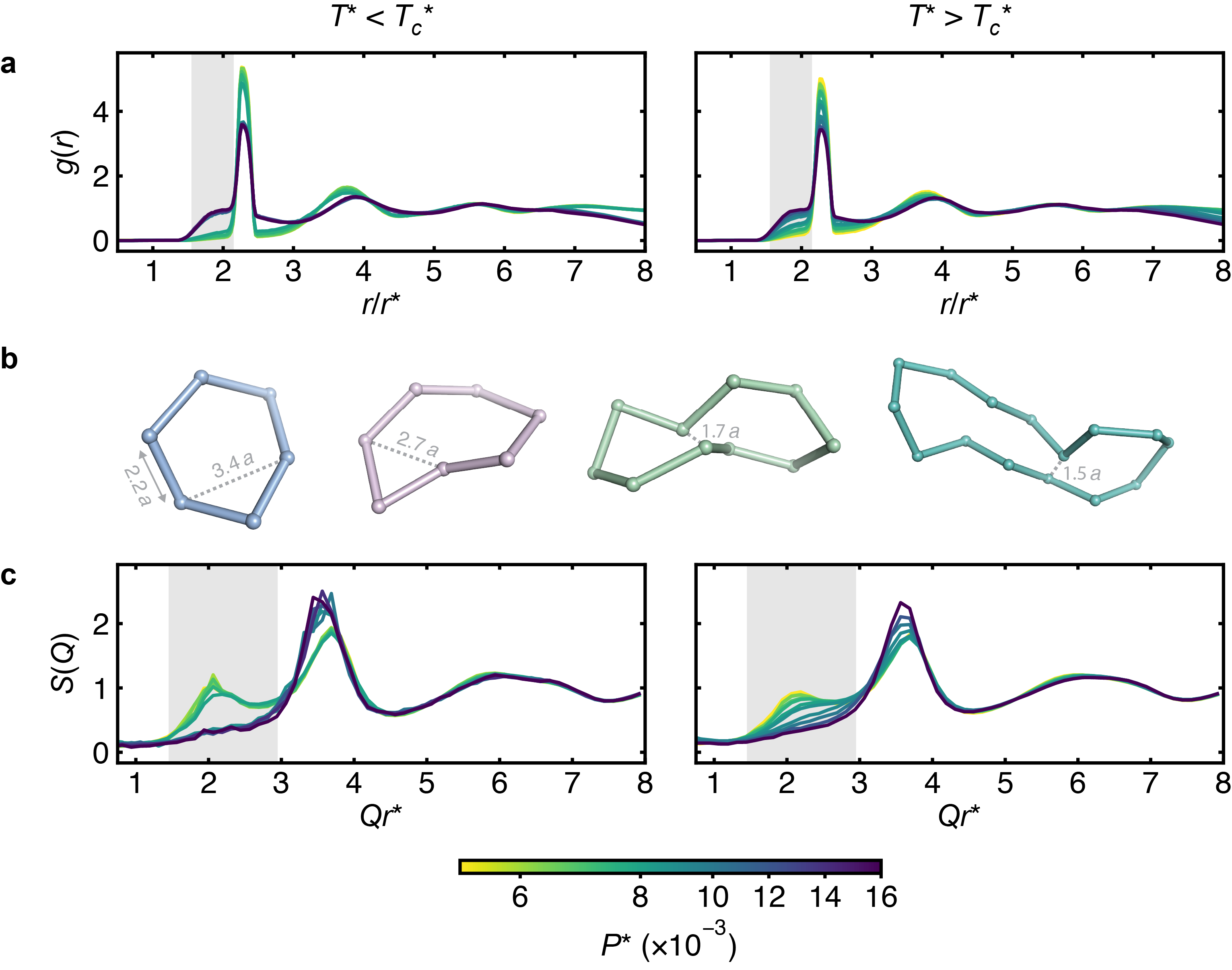}
	\end{center}
	\caption{\label{fig7} Radial distribution functions, ring geometries, and structure factors of the colloidal networks. (a) Pressure evolution of the radial distribution functions at low and high temperatures. The shaded region highlights reduced distances at which the close approach of non-bonded node pairs occurs.
    (b) Representative rings of increasing size (6-, 7-, 10- and 15-membered) taken from an HDL network. Dotted lines mark the shortest separation between non-bonded nodes in each ring; for compariosn, $2.2a$ is among the shortest bonded separations in the system. As ring size increases, the ring is able to fold back on itself, bringing non-bonded nodes closer together in space than bonded neighbours. (c) Pressure dependence of the corresponding structure factors. The low- and high-temperature series are shown in the left and right panels, respectively. The shaded regions highlight the low-$r$ and low-$Q$ features that distinguish LDL and HDL.}
\end{figure}

Consequently, we determined the radial distribution functions for the centres of the tetrahedral clusters from the Monte Carlo configurations sampled along the two isotherms; the results are shown in Fig.~\ref{fig7}(a). In these functions, the well-defined first peak at a reduced distance just above $r^\ast=r/a=2$ corresponds to the expected separation for connected nearest-neighbour nodes.
At low temperatures and pressures, the first peak is isolated in the radial distribution function until additional smaller peaks appear for distances $r^\ast>\ \sim3$. But as the pressure is increased and the topological transition is traversed, a broad hump does indeed emerge at a smaller separation: this additional feature has the same form as noted in Ref.~\cite{Neophytou2024}, and arises from pairs of nodes that are close in real space but are not directly connected in the network [Fig.~\ref{fig7}(b)]. Whereas the emergence of the hump occurs suddenly as pressure is increased at low temperatures, the same feature appears in a much smoother manner for the higher-temperature case.

Changes in the radial distribution function necessarily imply changes in the structure factor, since the two functions are related by the Fourier transformation. Hence, we calculated the structure factors from the same data sets; our results are shown in Fig.~\ref{fig7}(c). In this representation, the topological transition is even more obvious, with the disappearance of a low-angle `pre-peak' that is strong for the LDL phase but absent from the HDL phase. Similar behaviour has been observed in the pressure-dependent response of other disordered network structures \cite{Hsu2008,Ciarella2016}, including that of amorphous phosphorus \cite{Zaug2008}. In that case (which employed experimental total scattering measurements), the existence of the pre-peak was associated with correlations in voids within the structure. The structural interpretation of its disappearance in that case was a model in which the covalent phosphorus network flexed to `fill-in' the voids originally present when pressure is successively applied. To the best of our knowledge, the possibility of a topological transition was not discussed at that time. In the colloidal system on which we focus here, the pre-peak may again be linked to void--void correlations, but its disappearance coincides closely with the emergence of network entanglement (itself functioning as a mechanism of filling void space).

\begin{figure}
	\begin{center}
		\includegraphics[width=\columnwidth]{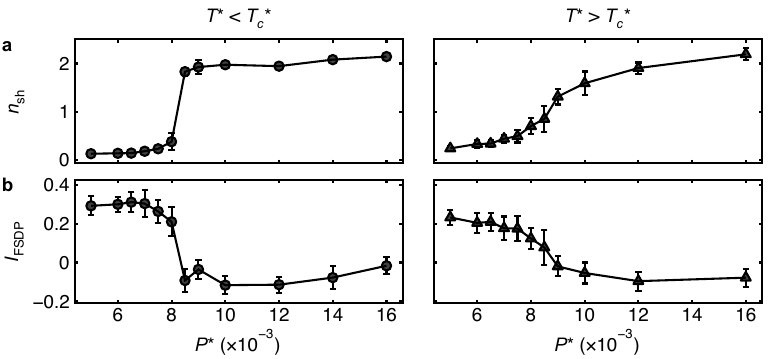}
	\end{center}
	\caption{\label{fig8}Order parameters extracted from the pair correlations, as a function of reduced pressure, below (left) and above (right) $T_{\rm c}^\ast$. (a) The number of nodes in close spatial proximity, $n_{\rm sh}=4\pi\rho\int r^2g(r)\,{\rm d}r$, integrated over the shaded window $1.55a\le r\le2.15a$ of Fig ~\ref{fig7}(a), where $\rho$ is the node number density. (b) The integrated intensity of the low-$Q$ pre-peak, $I_{\rm FSDP}=\int[S(Q)-S_{\rm b}(Q)]\,{\rm d}Q$, over the shaded window $1.45\le Qr^\ast\le2.95$ of Fig.~\ref{fig7}(c), where $S_{\rm b}(Q)$ is the straight line joining $S(Q)$ at the window edges. Error bars denote the standard deviation across independent configurations.}
\end{figure}

In Fig.~\ref{fig8}, we extract from our radial distribution function and structure factor calculations a pair of effective order parameters, based on the relative intensities of the features discussed above. The definitions of these order parameters are given in the Figure legend. Once again, we find that the pressure dependence of these order parameters tracks closely that derived directly from the topological measure of linking number as shown in Fig.~\ref{fig2}. By contrast to the ring-statistics measures discussed above, it is less straightforward to assert that this correspondence in order parameters is due to a common topological origin, as the (dis)continuous variation in configurational density across the transition will necessarily correlate with changes in peak positions and intensities in both radial distribution function and structure factor. Interestingly, the trends in peak positions that one might normally expect to follow density changes do so in an entirely counterintuitive way. Take the peak near $r^\ast=4$ in the pair distribution function, for example, which shifts to higher distances as pressure is applied. Likewise in reciprocal space, where the scattering feature near $Qr^\ast=6$ progresses to lower angle (larger distances) at high pressures. Such variations cannot be explained by uniform densification alone and instead reflect a substantial reorganisation of the network structure across the transition. One way or the other, what we have shown is that the scattering functions also have sensitivity to the topological transition---at least in this particular colloidal model.

\subsection{Comparison to other tetrahedral CRNs}

\begin{figure}
	\begin{center}
		\includegraphics{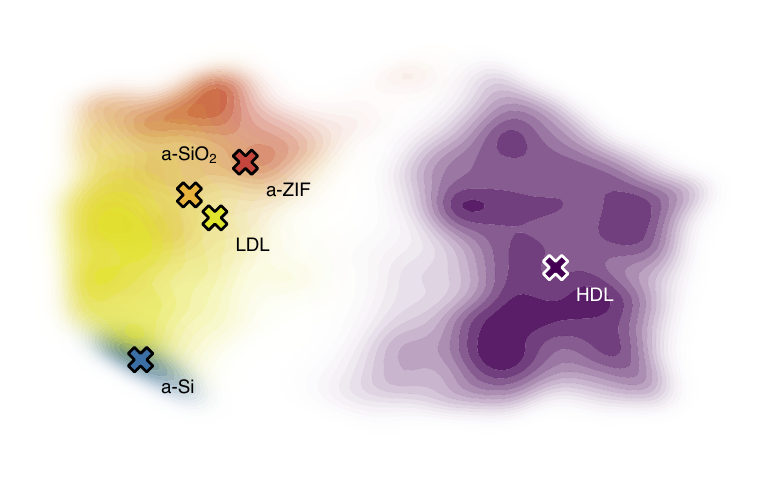}
	\end{center}
	\caption{\label{fig9}UMAP embedding of the topological space spanned by the colloidal LDL and HDL networks together with the amorphous tetrahedral networks a-Si, a-SiO$_2$ and a-ZIF of Refs.~\cite{Deringer2018,Erhard2022,Nicholas2025}. Shaded distributions show the density of node-level CARVS vectors for each system, and crosses mark the projections of the corresponding network-averaged CARVS vectors $\bar{\mathbf c}$. The embedding is the same as in Fig.~\ref{fig5}, fitted to the node-level CARVS vectors of all amorphous networks considered in this study. The LDL network lies close to a-SiO$_2$, within the region of topological space occupied by the conventional amorphous tetrahedral networks, whereas the HDL network occupies a distinct region not populated by any of them. [LDL and HDL configurations are taken at $T^\ast=0.1050$, $P^\ast=0.0050$ and $0.0160$, respectively.]}
\end{figure}

The final part of our (re)analysis of the colloidal water model places the LDL and HDL structures in the context of other disordered tetrahedral networks using their CARVS representations. Here we are building on the related earlier comparison between amorphous inorganic and metal--organic networks made first in Ref.~\cite{Nicholas2025}. In Fig.~\ref{fig9}, we return to the common UMAP embedding introduced in Fig.~\ref{fig5}, and now show the distributions of node environments found in models of three canonical systems: amorphous silicon \cite{Deringer2018}, amorphous silica \cite{Erhard2022}, and an amorphous zeolitic imidazolate framework \cite{Nicholas2025}. Because the UMAP embedding was fitted using node-level CARVS vectors from all five systems, these distributions can be compared directly within the same representation of CARVS space.

What we find is that these canonical systems all exhibit node topologies that are much more similar to the LDL state than the HDL state. This is perhaps to be expected since the LDL and HDL configurations occupy distinct regions corresponding to predominantly unentangled and strongly entangled network structures, respectively. Following the approach developed in Ref.~\cite{Nicholas2025}, Table~\ref{table1} summarises the CARVS vectors for each of the five systems involved. A quantitative comparison using Euclidean distances between the corresponding average CARVS vectors in the original high-dimensional space shows that the LDL state lies closest to amorphous silica. Given the proposed relevance of the colloidal water model to (real) liquid water, it is perhaps interesting to recall that one of the earliest models for the structure of liquid water, due to Bernal, was based on a silica-like disordered tetrahedral network \cite{Bernal1933,Finney2013}. More generally, the sequence from amorphous silicon through amorphous silica to the zeolitic imidazolate framework is accompanied by a progressive increase in both the size and diversity of the rings represented in the CARVS vectors. The amorphous ZIF therefore extends furthest towards the large-ring character that distinguishes the HDL state, although it remains clearly separated from the HDL distribution in CARVS space. This suggests that tetrahedral framework materials with suitably extended linkers may provide promising systems to observe non-trivial topological features, leading to entanglement.

\begin{table}[b]
\caption{\label{table1} CARVS for disordered tetrahedral networks.}
\begin{tabular}{@{}lll}
\br
System&CARVS&Ref.\\
\mr
{\it a}-Si&$\{5_{2.0}.6_{5.2}.7_{4.1}\}_{\sigma=3.5}$&\cite{Deringer2018}\\
{\it a}-SiO$_2$&$\{5_{1.6}.6_{2.3}.7_{2.6}.8_{2.1}.9_{1.5}\}_{\sigma=4.2}$&\cite{Erhard2022}\\
{\it a}-Zn(im)$_2$&$\{5_{1.3}.6_{2.1}.7_{3.0}.8_{2.9}.9_{2.9}.10_{1.8}.11_{1.5}\}_{\sigma=5.8}$&\cite{Nicholas2025}\\
LDL&$\{5_{1.7}.6_{2.6}.7_{2.6}.8_{1.5}.9_{1.2}\}_{\sigma = 3.8}$&This work\\
HDL&$\{5_{1.0}.6_{3.1}.7_{2.1}.8_{1.1}.9_{1.9}.10_{2.6}.11_{3.8}.12_{3.7}.13_{3.8}.14_{2.1}.15_{1.5}\}_{\sigma = 8.4}$&This work\\
\br
\end{tabular}
\end{table}

\section{Concluding Remarks}

In summary, our study has shown how order parameters based directly on ring statistics, on the one hand, and pairwise correlation functions, on the other hand, provide clear signatures of the topological transition in a colloidal tetrahedral network model. This sensitivity emerges despite the absence of explicit linking information in either class of descriptor, and reflects the structural reorganisation that accompanies the change in network entanglement. The character of the topological transition (\emph{viz.} discontinuous or continuous) is preserved in this indirect mapping, suggesting that scattering experiments and their interpretation \emph{via} advanced modelling methodologies may allow experimentally-driven characterisation of topological transitions in suitably-chosen physical systems. The CARVS analysis that we have performed shows a clear distinction between HDL and LDL topologies, and helps position other amorphous tetrahedral networks in the context of entanglement transitions. This analysis suggests that framework materials based on suitably extended linkers may provide promising candidates in which to search for such transitions---a direction we intend to pursue in the near future.

\ack
	This research was supported financially by the European Research Council (Advanced Grant 788144), the UKRI through Frontier Research Grant EP/Z534031/1, and the Royal Society through the Faraday Discovery Fellowships Fund, provided by DSIT. T.C.N. acknowledges support from the Research Foundation-Flanders (FWO Junior Postdoctoral Fellowship, grant no. 1264226N). The authors acknowledge open-access data accompanying Ref.~\cite{Neophytou2022} and insightful discussions with Andreas Neophytou. D.C. acknowledges support from the Leverhulme Research Fellowship (RF-2024-626) and thanks Francesco Sciortino for stimulating discussions.
	
\clearpage

\section*{References}
\bibliographystyle{iopart-num}
\bibliography{ref}

\end{document}